\documentclass[final,5p,times]{elsarticle}
\usepackage{amssymb}
\usepackage{graphicx}
\usepackage{array}
\usepackage{color}
\usepackage{url}
\biboptions{numbers,sort&compress}

\journal{Nuclear Instruments and Methods in Physics Research B}
\newcommand{\tsups}[1]{\textsuperscript{#1}}
\begin{document}

\begin{frontmatter}



  \title{Expanding RIB Capabilities at the Cyclotron Institute: \tsups{3}He-LIG
    production with an Isobar Separator LSTAR}


\author[CI,PHYS]{D. Melconian}
\cortext[cor1]{Corresponding author.}
\ead{dmelconian@tamu.edu}
\author[ND]{G.P.A. Berg}
\author[CI]{P.D. Shidling}
\author[ND]{M. Couder}
\author[ND]{M. Brodeur}
\author[CI]{G. Chubarian}
\author[CI]{V.E. Iacob}
\author[CI]{J. Klimo}
\author[CI]{G. Tabacaru}
\affiliation[CI]{Cyclotron Institute, Texas A\&M University, College Station, TX USA 77843-3366}
%organization={Cyclotron Institute, Texas A\&M University},
%            city={College Station},
%            postcode={77843}, 
%            state={TX},
%            country={USA}}
\affiliation[PHYS]{Department of Physics \& Astronomy, Texas A\&M University, College Station, TX USA 77843-4242}
  %organization={Department of Physics \& Astronomy, Texas A\&M University},
%            city={College Station},
%            postcode={77843-4242}, 
%            state={TX},
%            country={USA}}
\affiliation[ND]{Department of Physics \& Astronomy, University of Notre Dame, Notre Dame, IN USA 46556}
%  organization={Department of Physics \& Astronomy, University of Notre Dame},
%            city={Notre Dame},
%            postcode={46556},
%            state={IN},
%            country={USA}}
\begin{abstract}
  A new \tsups{3}He-driven IGISOL production station and mass separator 
  have been designed to produce neutron-deficient low-mass isotopes at 
  the Cyclotron Institute for the TAMUTRAP facility.  The LSTAR design has a mass resolution 
%  is able to completely separate contaminant masses with 
  $M/\Delta M\geq3,000$ to reject contaminants with $>95\%$ efficiency.
  %  An overview of the He-LIG and LSTAR systems and their expected performance
  %  will be presented, largely within the context of the TAMUTRAP science
  %  program.
\end{abstract}



\begin{keyword}
High-resolution isobar separator \sep Light-ion guide \sep Charged-particle spectrometers
%\PACS 29.30.Aj

\end{keyword}

\end{frontmatter}


\section{Introduction}\label{sec:intro}
The primary goal of a new facility, TAMUTRAP, at the Cyclotron Institute,
Texas A\&M University, is to look for physics beyond the standard model 
by searching for possible scalar or tensor currents contributing to the 
weak interaction in nuclear $\beta$ 
decay~\cite{JTW,falkowski,shidling-Hyper}.
In particular, it will enable the measurement of the $\beta$-$\nu$ correlation parameter
%$\tilde{a}_{\beta\nu}=a_{\beta\nu}/(1+b\frac{m}{\langle E\rangle})$, 
in several $T=2$ and $T=3/2$ superallowed $\beta$-delayed proton
emitters initially confined in a novel and unique cylindrical Penning 
trap~\cite{mehlman}.  This trap has been designed to be very large 
(180-mm inner diameter) so that
$\beta$-delayed protons of up to 4.25~MeV energy are fully contained
radially by the 7-T field of the magnet. We have commissioned the 
TAMUTRAP facility by demonstrating the ability to
perform precise mass measurements using offline ion sources. Once the radioactive ion beam (RIB) is
successfully produced and transported to TAMUTRAP, we will be uniquely
suited to observe the $\beta$-delayed proton decays of \tsups{20,21}Mg,
\tsups{24,25}Si, \tsups{28,29}S, \tsups{32,33}Ar and \tsups{36,37}Ca with
$4\pi$ collection of the $\beta$s and delayed protons.  An overview of the TAMUTRAP facility and scientific program is described in Ref.~\cite{shidling-IntJ21}.

%\begin{figure}\centering
%  \includegraphics[width=0.95\linewidth]{CI-TAMUTRAP-LSTAR-HeLIG.png}
%  \caption{The He-LIG production station and LSTAR separator upgrade planned at the Cyclotron Institute.\label{fig:chart-of-nuclides}}
%\end{figure}
Here we describe our plans for producing the proton-rich beams for 
TAMUTRAP (the ``He-LIG'' station) and the separator (``LSTAR'') 
designed to purify and transport the beams to the TAMUTRAP facility.

\section{RIB Production: \tsups{3}He-LIG}
The K150 cyclotron has been recommissioned at the Cyclotron Institute 
(CI) at Texas A\&M to produce these radioactive ion beams using the 
IGISOL (Ion Guide Isotope Separator On-Line) 
technique~\cite{moore-HypInt14}.  
In this technique the nuclei of interest are 
produced by a primary beam of protons on a thin 
target. The low-energy fusion-evaporation reaction products enter a 
small chamber ($\sim$1~cm$^3$) filled with ultra-high purity helium gas. 
The products are slowed down in the gas and transported by a gas flow 
out of the gas cell and injected through a differentially-pumped
electrode system to the high-vacuum section of an isotope separator for 
further acceleration and separation. The IGISOL technique extracts ions 
very quickly ($\approx$1~ms) allowing for the extraction of short-lived 
nuclei with typically $\approx$1\% efficiency~\cite{aysto-NuclPhysA}. 

The current $p$-LIG effort at the CI uses up to 40-MeV protons for $(p,n)$ 
reactions, charge-breeds the extracted reaction products in a ECRIS before 
injecting into the CI's K500 superconducting cyclotron for 
re-acceleration~\cite{tabacaru}.  In order to produce the neutron-deficient 
RIBs of interest for TAMUTRAP via ($p,2n$) and ($p,3n$) reactions, adequate production requires using $(10-22)$~MeV/u
\tsups{3}He as the driver for the LIG station rather than protons; thus 
we are developing a new ``He-LIG'' system which will share a gas-cell 
chamber with the current $p$-LIG, but use separate gas-cells and extract 
RIB in the opposite direction to a new beamline.

\begin{figure}\centering
    \includegraphics[width=0.99\linewidth]{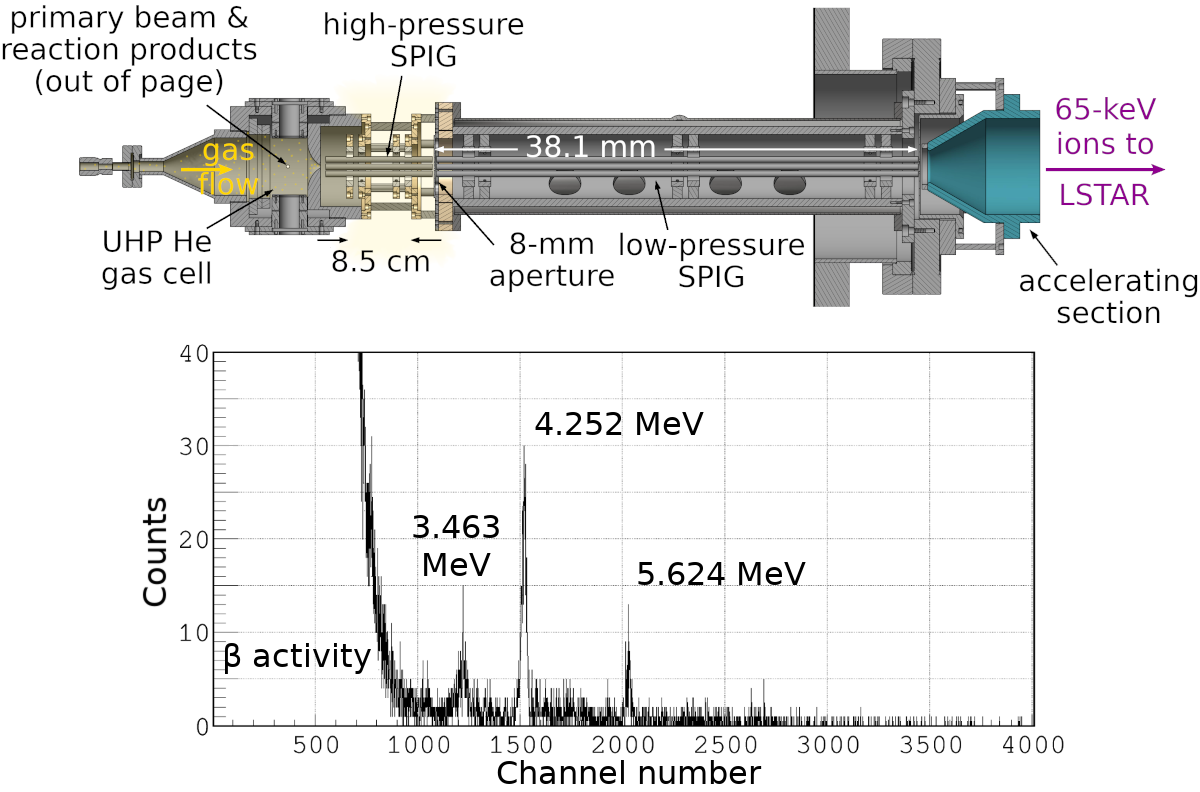}
    \caption{The He-LIG gas cell and SPIG system (top) and observed decay spectrum with our test cell (bottom). The K150 $^3$He primary 
    beam reacts with a thin target mounted in front of the gas cell (not 
    shown, out of the page). The reaction products are stopped in the gas 
    cell are quickly transported to the two-stage SPIG transport system 
    by the high flow of helium gas. The low energy part of the decay product spectrum is dominated by $\beta$s, but the proton peaks from \tsups{25}Si are clearly seen.\label{fig:si25}}
\end{figure}

%The high-intensity primary beam of \tsups{3}He from the K150 cyclotron will
%impinge on a heavy target with the reaction products collected and extracted
%using the light-ion guide (LIG) technique~\cite{moore-HypInt14}. Following this we have designed
%LSTAR, a compact, high-resolution isobar separator to purify the RIBs in
%order to prevent overloading TAMUTRAP's RFQ cooler and buncher with
%contaminants.

We have tested a prototype gas cell, depicted in the top panel of Fig.~\ref{fig:si25}, 
using a \tsups{nat}Mg target 
and observed the three prominent proton branches following the $\beta$ 
decay of $^{25}$Si (see the lower panel of Fig.~\ref{fig:si25}). The efficiency of the system 
was about $1-2$ orders of magnitude below other LIG systems (e.g.\ at 
Jyv\"askyl\"a), but our updated design addresses many expected deficiencies of 
the prototype and we expect to reach efficiencies of $\geq$1\%. We have 
designed a new chamber which will allow for precise alignment of both $p$-
and He-LIG stations as well as quick and simple changeover procedures 
between the two systems. Once the new He-LIG system is assembled, we will 
characterize its efficiency and emittance using the pepper-pot technique.

\section{RIB Purification: LSTAR}
\begin{figure}\centering
    \includegraphics[width=0.99\linewidth]{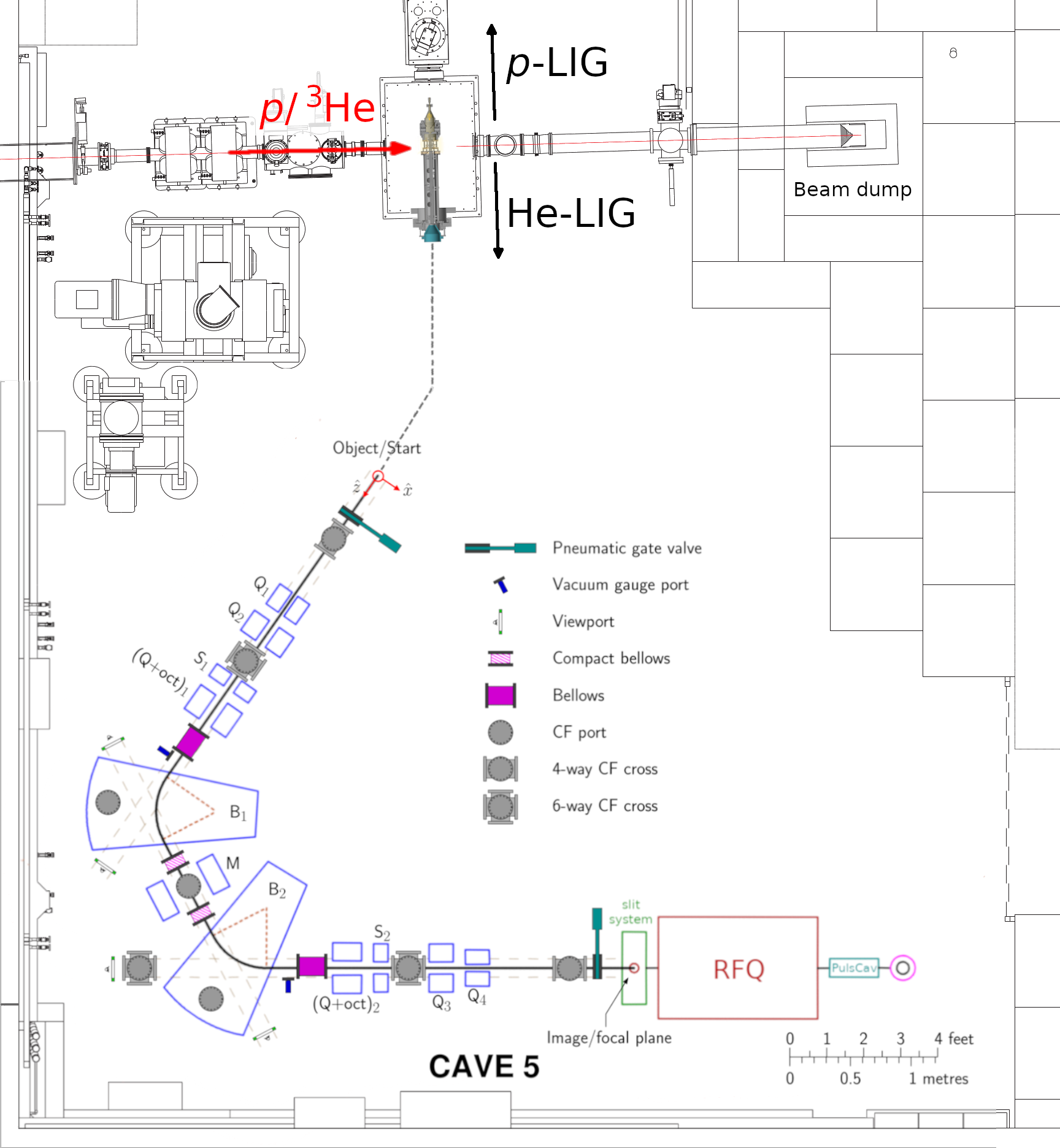}
    \caption{Top view of Cave 5 showing the
      planned layout of LSTAR\@. RIB produced in the He-LIG will be transported out of the $p$/He-LIG chamber with SPIGs and accelerated to 65~keV to the object of LSTAR\@. The RFQ placed at the focal plane of LSTAR will collect the purified RIB, allowing easy transport of cooled, bunched beams to the TAMUTRAP facility.
    \label{fig:layout}}
\end{figure}

The isotopes of interest for the TAMUTRAP program are relatively far from stability, so the production rates of nearby contaminants are a concern because too much beam can overload the RFQ cooler and buncher. We therefore require an isobar separator capable of purifying the produced RIB %\@. Mass resolutions of $M/\Delta M\geq3,000$ are needed to purify our most
well enough to purify the most
difficult case, separating \tsups{36}Ca from \tsups{36}K\@.  The most significant design requirements for the separator are:
\begin{enumerate}
\item A mass resolving power of $M/\Delta M\geq3,000$ for $6\leq A\leq 50$,
\item High acceptance and transmission ($>95\%$),
\item No energy compensation (meaning no electric dispersive elements),
\item Purely electrostatic focusing elements (so that settings are independent of mass), and
\item The separator must be compact to fit in the existing space available in Cave 5 of the Cyclotron Institute (see Fig.~\ref{fig:layout}) and have the focus be at the entrance to TAMUTRAP's RFQ cooler and buncher.
\end{enumerate}

\begin{figure}\centering
    \includegraphics[width=0.99\linewidth]{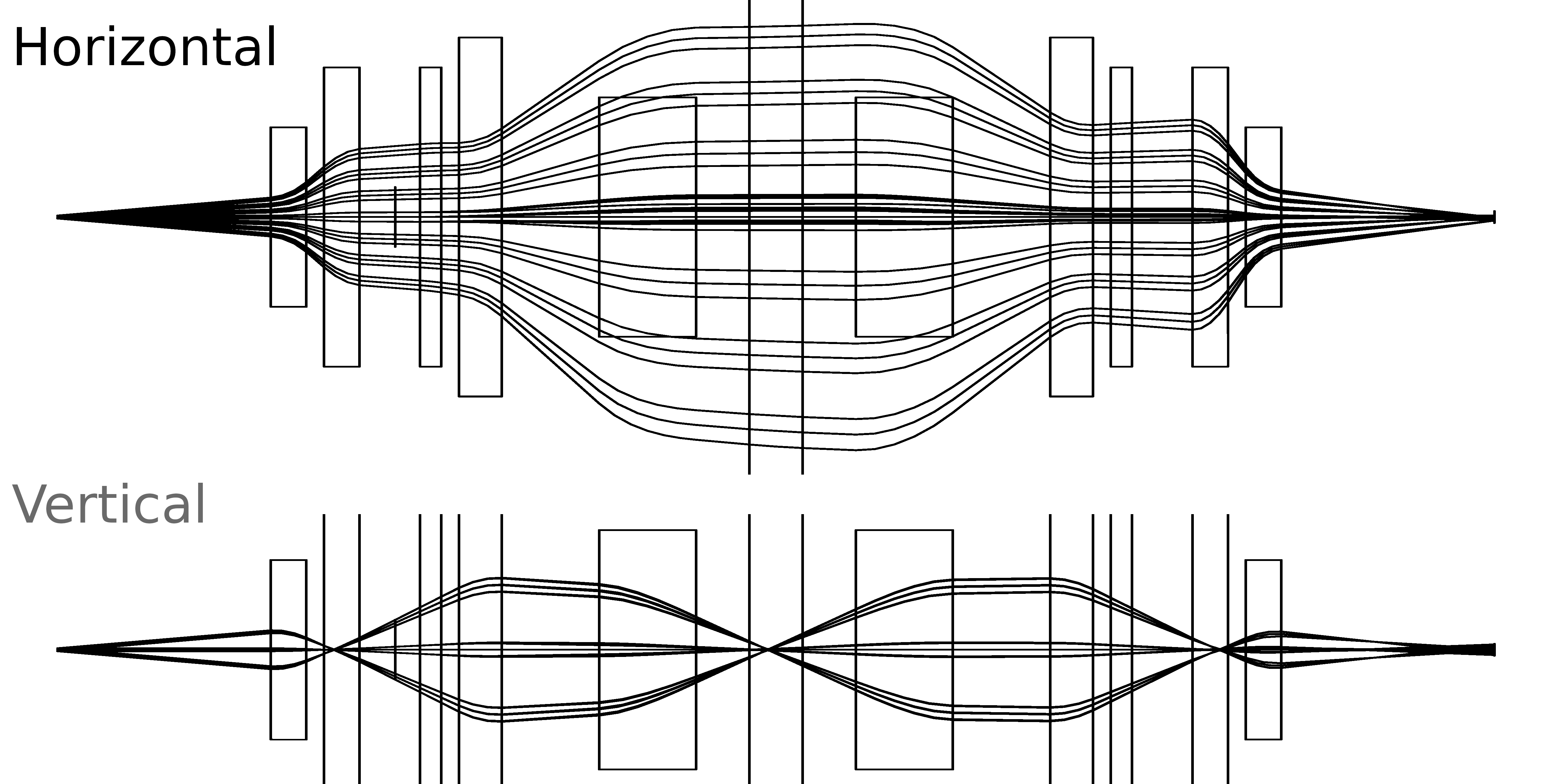}
    \caption{Dispersive (horizontal) and non-dispersive (vertical) ion optics of the LSTAR design.\label{fig:COSY-rays}}
\end{figure}

Our design, the Light-ion guide Separator for 
Texas A\&M's Radioactive ion beams (LSTAR), was inspired by the concept 
of CARIBU's separator at ANL~\cite{davids-NIM} since it shares many of 
the design specifications.  The elements of our design and how it will 
couple to the He-LIG are schematically shown in Fig.~\ref{fig:layout}.  
The primary 
$p$ or \tsups{3}He beam from the K150 enters from the left and the RIBs are 
produced in the dual $p$/He-LIG chamber; when running in He-LIG mode, the 
RIB will be directed down and turned $30^\circ$ via a curved sextupole ion guide (SPIG) and then 
accelerated to 65~keV to the object of LSTAR\@.  
After a 1.2~m long drift, the beam is tuned by several electrostatic 
elements, bent horizontally by two $62.5^\circ$ dipole magnets B$_1$ and 
B$_2$, with an electrostatic multipole, M, between the dipoles, and focused 
by several electrostatic quadrupoles onto the focal plane. Two 
electrostatic hexapoles, S$_1$ and S$_2$, are included in the combined 
function multipoles (Q+oct)$_{1,2}$.  The layout is symmetric about
the central electrostatic multipole M to reduce higher-order aberrations. 
This ion optics for this system, shown with sample ray traces in 
Fig.~\ref{fig:COSY-rays}, has been optimized to 3\tsups{rd} order and verified to 7\tsups{th} order using the
program COSY INFINITY~\cite{cosy}.  The technical design of LSTAR will be 
described in detail in a forthcoming publication~\cite{LSTAR-2bpublished}.

\begin{figure}\centering
    \includegraphics[width=0.90\linewidth]{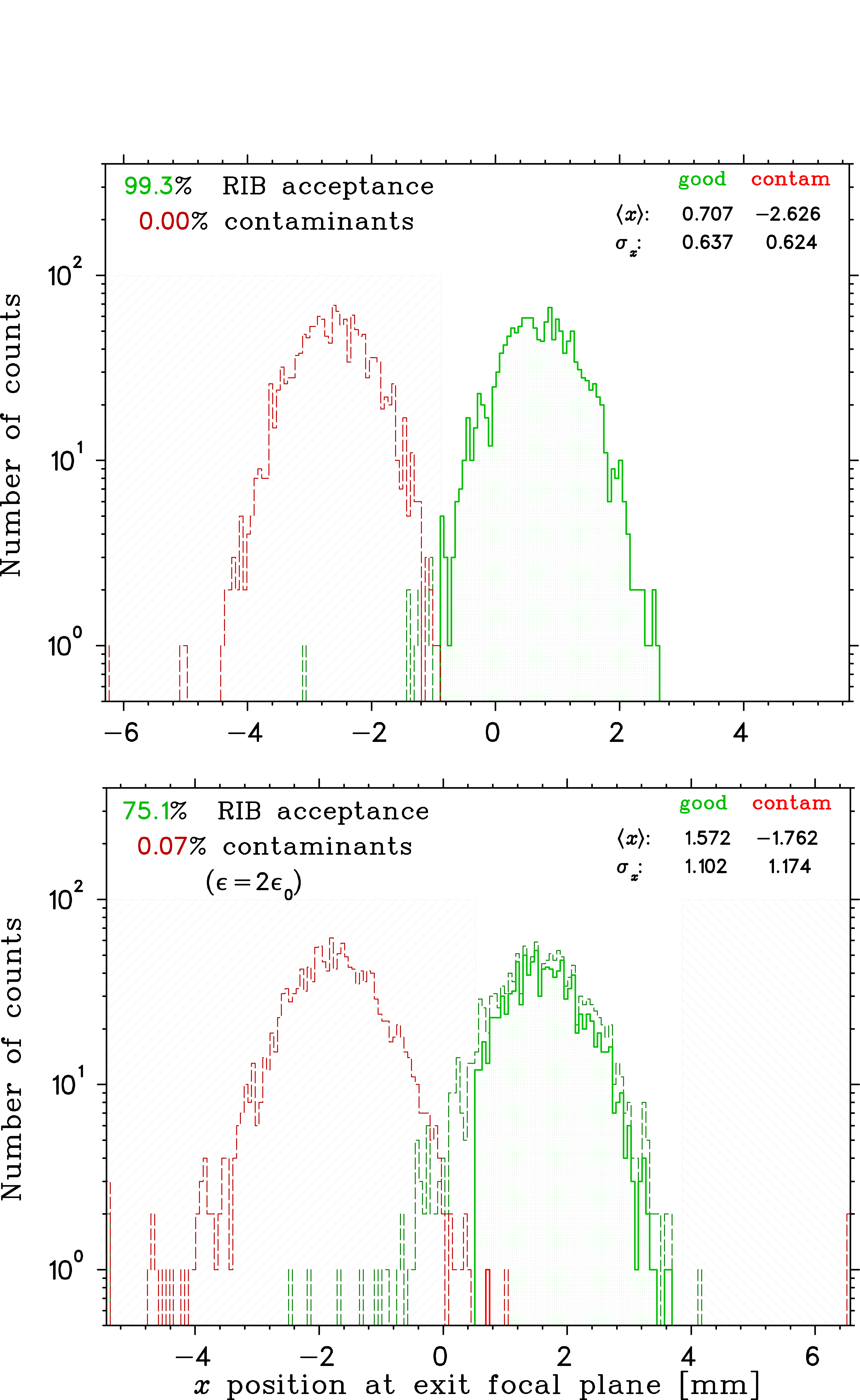}
    \caption{Performance of LSTAR using the initial emittance and energy spread as simulated by SIMION.  Top: distribution of good (green) and $\Delta M/M=3,000$ contaminant (red) ions at the focal plane of LSTAR\@. Bottom: same, but with twice the emittance estimated using SIMION. Complete separation of the contaminants is still achievable with 75\% acceptance of the species of interest..\label{fig:performance}}
\end{figure}

We have performed detailed studies of the expected performance of 
LSTAR based on the ion optics calculated in COSY\@. We start with the 
distribution of ion positions and velocities from the end of the SPIG of the He-LIG system as 
simulated by SIMION~\cite{simion}. Using these rays as input, we accelerated the ions to 65~keV and traced them
through LSTAR using COSY.  The same initial rays but with a mass % as well as a nearby contaminant with a mass 
difference $\Delta M/M=3,000$ were also traced to determine how well contaminants will be vetoed. The studies included the 
$0.65\pi\,\mathrm{mm}\,\mathrm{mrad}$ emittance and 3.33~eV energy 
spread predicted by SIMION. Position and angular misalignments of the 
elements of LSTAR were randomly varied as Gaussians of widths 0.25~mm and 0.25~mrad, respectively, and the results averaged to 
determine the tolerances required. Figure~\ref{fig:performance} shows the distribution of ions at the focal plane of LSTAR (top) and how broadened the distributions are if the initial emittance is twice what SIMION predicts (bottom).  Similar results are seen when other tolerances ($\Delta E$ and positions/angles of elements) are similarly broken.

In summary, LSTAR is expected to have a $\geq95\%$ efficiency 
for the ions of interest while completely removing all contaminants if
position and angular misalignments are $\leq$0.25~mm and
$\leq$0.25~mrad, respectively, and SIMION estimates of the emittance and energy spread are realistic. 
%Even if the emittance is $2\times$ worse than predicted, LSTAR will 
%still have $\geq$75\% transmission efficiency and a $\geq$99.8\% veto efficiency for 
%contaminants, as seen in the right plot of Fig.~\ref{fig:performance}.

\section{Outlook}
A new beamline for the production and purification of proton-rich 
beams, which will enable TAMUTRAP's studies of 
$\beta$ decay to test the standard model has been described.  Detailed studies of the 
design indicate the separator will completely remove contaminants with 
masses $M/\Delta M\geq3,000$ with $>95\%$ transmission.  Longer term 
future prospects include using TAMUTRAP for astrophysical studies with the Rogachev group, and 
injecting the cooled, bunched RIB into the K500 after charge-breeding with an EBIT.

\section{Acknowledgements}
This work was supported by the U.S. Department of Energy, Ofﬁce of Science, under Award No. DEFG02-93ER40773, and by the Department of Energy, National Nuclear Security Administration under Award No. DE-NA0003841 (CENTAUR). 
G.P.A.B., M.B., and M.C. acknowledge support of the National Science Foundation under Grant PHY-2011890.





\end{document}